\documentclass[pra,
twocolumn,
 amsmath,amssymb,
 aps,
floatfix]{revtex4-2}

\usepackage[utf8]{inputenc}
\usepackage{color}

\usepackage{lmodern}
\usepackage{amsmath}
\usepackage{amssymb}
\usepackage{mathtools}
\usepackage{rotating}

\usepackage[pagebackref=false,colorlinks,linkcolor=blue,filecolor=green,urlcolor=blue ,citecolor=blue]{hyperref}
\usepackage[english]{babel} 

\usepackage[tight,nice]{units}

\DeclareMathAlphabet{\bi}{OML}{cmm}{b}{it}

\newcommand{\rmi}{\mathrm{i}}

\newcommand{\imagi}{\rmi}

\newcommand{\emptysite}{\!\cdot\!}

\newcommand{\eulere}{\mathrm{e}}

\newcommand{\beq}{\begin{equation}}
\newcommand{\eeq}{\end{equation}}

\newcommand{\makegreen}[1]{{ #1}}  

\definecolor{lightyellow}{cmyk}{0.0,0.0,0.2,0.0}
\definecolor{orange}{rgb}{0.9,0.6,0.0}
\definecolor{gold}{rgb}{0.8,0.6,0.0}
\definecolor{lightcyan}{cmyk}{0.2,0.0,0.0,0.0}
\definecolor{cyan}{cmyk}{1.0,0.0,0.0,0.0}
\definecolor{lightlila}{cmyk}{0.0,0.1,0.0,0.0}
\definecolor{lila}{cmyk}{0.0,1.0,0.0,0.2}
\definecolor{darklila}{cmyk}{0.0,1.0,0.0,0.4}
\definecolor{brown}{cmyk}{0.0,0.3,0.3,0.3}
\definecolor{transparent}{cmyk}{0.0,0.0,0.0,0.0}
\definecolor{lilapluscyan}{cmyk}{1.0,1.0,0.0,0.2}
\definecolor{bauer}{rgb}{0.04,0.31,0.63}
\definecolor{darkgreen}{rgb}{0.0,0.5,0}
\definecolor{darkyellow}{rgb}{0.8,0.7,0.0}
\definecolor{darkred}{rgb}{0.5,0.0,0}

\begin{document}

\title{Harmonic generation with topological edge states and electron-electron interaction}

\author{S.\ Pooyan} 
\author{D.\ Bauer}
\affiliation{Institute of Physics,  Rostock University, 18051 Rostock, Germany}

\date{\today}

\begin{abstract} 
It has been found previously that the presence or absence of topological edge states in the Su-Schrieffer-Heeger (SSH) model has a huge impact on harmonic generation spectra. More specifically, the yield of harmonics for harmonic orders that correspond to photon energies below the band gap is many orders of magnitude different in the trivial and topological phase. It is shown in this work that this effect is still present if  electron-electron interaction is taken into account, i.e., if a Hubbard term is added to the SSH Hamiltonian. To that end, finite SSH-Hubbard chains at half filling are considered that are short enough to be accessible to exact diagonalization but already showing edge states in the topological phase. We show that the huge difference in the harmonic yield between the trivial and the topological phase can be reproduced with few-level models employing only the many-body ground state and a few excited many-body states.    
\end{abstract}

\maketitle

\pagebreak

\section{Introduction}
Condensed-matter high-harmonic spectroscopy stands as a burgeoning field in strong-field science with ultrashort laser pulses. This technique facilitates the all-optical, ultrafast examination of both structural and dynamic properties of materials \cite{bauer2018high, jurss2019high, Ghimire2010, Schubert2014, Vampa2015, Hohenleutner2015, Luu2015, Ndabashimiye2016, Langer2017, TancogneDejean2017, You2017, Zhang2018, Vampa2018, Baudisch2018, Garg2018}. Research efforts in condensed matter physics  have focused on topological phases, with a specific emphasis on areas such as topological insulators \cite{Hasan2010, Ando2013, Chiu2016, Moore2010}, topological superconductors \cite{Qi2011, Viyuela2018}, photonic topological insulators \cite{Rechtsman2013, Sttzer2018, Kruk2018}, and topological circuits \cite{Ningyuan2015, Albert2015}. It was only recently that both theoretical investigations \cite{bauer2018high, jurss2019high,Drueke2019, Moos2020, KoochakiKelardeh2017} and experimental studies \cite{Luu2018, Reimann2018} initiated an exploration into the physics occurring at the intersection of strong-field high-harmonic generation (HHG) and topological condensed matter physics. Although clear spectral features in HHG from bulk that unambiguously would pinpoint topological properties are not yet found \cite{Neufeld2023} it is worthwhile to examine the explicit effects of topological edge states on HHG, which are not accessible to bulk simulations. 

HHG in interacting systems have garnered increasing interest. Already 20 years ago,   Baldea {\em et al.} \cite{baldea2004high} explored HHG in a hexagonal quantum-dot ring using the Pariser-Parr-Pople model, incorporating Hubbard-onsite and next-nearest neighbor interactions. Their findings indicated that such a system could produce higher harmonic intensities compared to natural benzene, highlighting the potential of engineered quantum structures in enhancing HHG efficiency. More recently, Silva {\em et al.}\  \cite{silva2018high} conducted a theoretical study on HHG spectroscopy of light-induced Mott transitions within the Fermi-Hubbard model. Their work demonstrated the feasibility of achieving high time resolution, which is crucial for studying ultrafast dynamics in correlated materials. Murakami {\em et al.} \cite{murakami2018high} employed the Hubbard model to describe HHG in a Mott insulator, including coupling to an environment. They focused on the roles of doublons and holons, providing insights into HHG in highly correlated materials. In another work, Murakami {\em et al.}\ \cite{murakami2021high} investigated HHG due to doublon-holon recombination. They developed a three-step model for doublon-holon pairs and showed that spin dynamics could significantly influence HHG efficiency, indicating the sensitivity of HHG to spin dynamics in Mott systems.  Tancogne-Dejean {\em et al.}\  \cite{tancogne2018ultrafast} investigated the changes in the Hubbard $U$ parameter under laser excitation using density functional theory simulations. They demonstrated how laser fields could be used to tune material properties and drive phase transitions, specifically in NiO. Lysne {\em et al.}\  \cite{lysne2020signatures} extended the analysis to more complex systems, incorporating dynamical mean-field theory to account for coupling to bosons or local two-orbital models with Hubbard interaction and Hund coupling. This approach allowed for a more nuanced understanding of HHG in correlated electron systems. Imai {\em et al.}\ \cite{imai2020highharmonic} explored HHG in a spinless-fermion model on a dimer lattice, which can be mapped to an Ising model at low energies. They interpreted HHG in terms of recombination of kink-antikink excitations, providing a novel perspective on the underlying mechanisms.  Avchyan {\em et al.}\ \cite{avchyan2022laser} used an extended Hubbard model with longer-range interactions treated within a phenomenological Hartree-Fock approximation to study two-color HHG in graphene quantum dots,  emphasizing the effects of interaction range and system geometry on HHG. Hansen and colleagues \cite{hansen2022correlation} examined the effect of electron correlation and finite system size on HHG using a one-band Fermi-Hubbard model. They found that lower-order harmonics are mainly influenced by finite-size effects, while higher orders are more affected by electron correlation. In a subsequent study \cite{hansen2022doping} they explored the impact of doping on HHG, analyzing various Hubbard interaction parameter values. They interpreted the results in terms of doublon-holon pairs, shedding light on how different filling levels affect HHG. Valmispild {\em et al.}\  \cite{valmispild2024sub-cycle} utilized sub-cycle spectroscopy to investigate highly correlated materials. They employed a non-equilibrium extension of dynamical mean-field theory and analyzed the one-particle Keldysh Green's function, demonstrating the potential of manipulating strongly-correlated materials with non-resonant light fields.

In Refs.\ \cite{bauer2018high, jurss2019high}, a huge difference in the harmonic spectra for the trivial and the topological phase of finite SSH chains was observed. The origin of that difference in the harmonics yield for photon energies below the (bulk) band gap could be traced back to the presence of two edge states of which only one is populated. This result was obtained both for a time-dependent density-functional theory (TDDFT) treatment \cite{bauer2018high} and with the simpler tight-binding description \cite{bauer2018high, jurss2019high} of SSH chains. In both papers very long chains could be treated because electron-electron (e-e) interaction was neglected in the tight-binding treatment in \cite{jurss2019high}, and the auxiliary Kohn-Sham particles of TDDFT used in \cite{bauer2018high}, by construction, interact through an effective potential only. 

In this work, we will incorporate a short-range Hubbard interaction and solve the model for the finite SSH-Hubbard system by exact numerical diagonalization. As a consequence, we have to restrict ourselves to rather short chains. However, we will see that clear precursors of the topological edge states known from long chains without e-e interaction appear already in short chains with (not too strong) interaction. We find that the huge difference in the harmonic yield between the trivial and the topological phase survives modest e-e interaction. For stronger e-e interaction, HHG becomes inefficient in both phases.   Moreover, we find that  most of the computational cost needs to be spent on the field-free electronic structure of the interacting electron system while the propagation in the laser field is very cheap. This is because the relevant many-electron states that take part in HHG are only a small fraction of the total number of states. In fact, the first harmonic plateau can be calculated accurately with surprisingly few many-body states. 

We want to emphasize that this is not a study of HHG in bulk but takes the presence of (many-body) edge or surface states self-consistently into account because  the laser will inevitably  ``see'' these states when impinging on the target. Whether these edge or surface  states can be called ``topological'' in the interacting case or not is an interesting question that goes well beyond the specific topic of this work.

The paper is structured as follows. In Sec.\ \ref{sec:model}, a brief introduction to the SSH model with Hubbard term and its coupling to an external field is provided.  Harmonic spectra for both topological and trivial cases for twelve sites at half filling are presented in  Sec.\ \ref{sec:results}. In Secs.\  \ref{sec:dep_on_u} and \ref{sec:few_states}, the influence of the Hubbard interaction on HHG is studied,  and the sufficiency of a few many-electron states for the time-propagation is demonstrated. In Sec.\  \ref{sec:twelve_many_body_states}, the dominant configurations in the lowest-lying, relevant  many-body states are discussed. Section \ref{sec:summary} concludes the work and provides an outlook.

\section{Model} \label{sec:model}
\begin{figure}
\includegraphics[width=1\columnwidth]{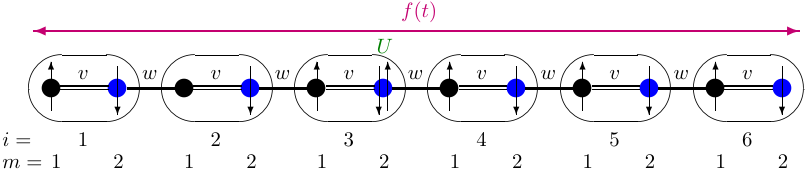}

\caption{The SSH-Hubbard chain for six unit cells $i=1, 2, \ldots, 6$. The intracell hopping amplitude is $-v$, and the intercell hopping amplitude is $-w$.  As an example, the possible positions of six spin-up and six spin-down electrons are indicated by up and down arrows. For that configuration, two electrons (of opposite spin) are located at site $i=3$, $m=2$ such that the Hubbard interaction $U$ contributes.  \label{fig:sshhubbardchainmain} }
\end{figure}

Consider a bi-partite chain with interacting spin-1/2 particles, e.g., electrons, on it. The corresponding tight-binding Hamiltonian reads  
\begin{align} H_0 &= -v \sum_{i\sigma} (c^\dagger_{i1\sigma} c_{i2\sigma} +  c^\dagger_{i2\sigma} c_{i1\sigma}) \nonumber \\
& \quad - w \sum_{i\sigma} (c^\dagger_{i+1,1\sigma} c_{i2\sigma} +  c^\dagger_{i2\sigma} c_{i+1,1\sigma}) \nonumber \\
& \quad + \frac{U}{2} \sum_{im\sigma\sigma'}  c^\dagger_{im\sigma} c^\dagger_{im\sigma'} c_{im\sigma'} c_{im\sigma}   . \label{eq:Hmain}
\end{align}
Here, we use the usual notation $\sigma=\uparrow,\downarrow$ for a spin, $i=1,2,3,\ldots $ labels the unit cells, and $m=1,2$ the sites within a unit cell. 
The operators $c_{im\sigma}$ and $c^\dagger_{im\sigma}$ are fermionic annihilation and creation operators, respectively, fulfilling the usual anti-commutator relations 
$ \{ c_{im\sigma}, c^\dagger_{i'm'\sigma'} \} = \delta_{ii'} \delta_{mm'} \delta_{\sigma\sigma'} $ and $  \{ c_{im\sigma}, c_{i'm'\sigma'} \} = \{ c^\dagger_{im\sigma}, c^\dagger_{i'm'\sigma'} \} = 0$. 
The first two sums over the hoppings $\propto v,w$ in \eqref{eq:Hmain} constitute the SSH model \cite{su1979solitons} while the third term $\propto U$ is the short-range Hubbard interaction.
Figure \ref{fig:sshhubbardchainmain} shows a sketch of a finite chain with six unit cells $i=1, 2, \ldots, 6$. In this work, we consider half-filling, i.e., as many electrons as lattice sites, half of the electrons spin-up, half spin-down.

We couple to an external driver in length gauge, e.g., a laser field in dipole approximation, polarized along the chain, by adding a time-dependent term 
\begin{align} H_\mathrm{field}(t) &= E(t) \sum_{im\sigma} x_{im} c^\dagger_{im\sigma}c_{im\sigma}  \label{eq:Hfield}
\end{align}
where $x_{im}$ is the position of the lattice site $(i,m)$ We considered equidistant lattice sites separated by one unit of length (i.e., a lattice constant $a=2$), with the origin in the center of the chain. The
 total Hamiltonian then reads
\beq H(t) = H_0 + H_\mathrm{field}(t) . \eeq

Choosing a particular basis for the many-electron Hilbert space, the field-free many-electron eigenstates $\mathbf{u}_j$, $j=0,1,2,\ldots n-1$ are calculated by solving the eigenvalue problem 
\beq \mathbf{H_0} \mathbf{u}_j = \varepsilon_j \mathbf{u}_j. \label{eq:H0matrixequ} \eeq
Here, $\mathbf{H_0}$ is the matrix representation of $H_0$ in the chosen basis, and $\varepsilon_0 \leq \varepsilon_1 \leq \varepsilon_2 \leq \ldots \leq \varepsilon_{n-1}$ are the eigenenergies of the many-electron eigenstates in ascending order. For the case of half-filling, the rank of the matrix $\mathbf{H_0}$  is $ n   =  \binom{N}{N/2}^2$. \makegreen{ The exact diagonalization approach is only feasible for not too large $N$ and depends on the computational resources available.  In the following, we show results for $N=12$ where $n=853776$. For vanishing interaction $U$, much larger systems can be treated \cite{jurss2019high}.   }

The matrix representation of $H_\mathrm{field}(t)$ factorizes in dipole approximation into $  \mathbf{H}_\mathrm{field}(t) = E(t)  \mathbf{h}$
where $\mathbf{h}$  is a time-independent matrix. Expanding the time-dependent state in field-free eigenstates,
\beq \boldsymbol{\psi}(t) = \sum_j b_j(t) \eulere^{-\imagi\varepsilon_j t} \mathbf{u}_j \label{eq:expansionineigenstatesImain}\eeq
including the field-free time evolution $\eulere^{-\imagi\varepsilon_j t}$,
we obtain
\begin{align}
 \imagi \dot b_k(t)     &=   E(t) \sum_j  \eulere^{-\imagi(\varepsilon_j-\varepsilon_k) t}  \bar T_{kj}b_j(t) \label{eq:eomforbsmain}
\end{align}
with
\beq \bar T_{kj} =  \mathbf{u}^\dagger_k   \mathbf{h} \mathbf{u}_j .\eeq
The expansion \eqref{eq:expansionineigenstatesImain} transforms from the Schr\"odinger picture to the interaction picture.  
 Transitions between states $k$ and $j$ will only take place if the transition-matrix element $T_{kj}=T_{jk}^*$ is non-vanishing.  

Collecting the time-dependent coefficients $b_k(t)$ in a column vector $\mathbf{b}(t)$, the time-dependent Schr\"odinger equation in the interaction picture    \eqref{eq:eomforbsmain} can be written in matrix form, 
\beq \imagi \dot{\mathbf{b}}(t) = \mathbf{M}(t) \mathbf{b}(t) \label{fig:eomforbvectormatrixformmain} \eeq
with the matrix elements of $\mathbf{M}(t)$ being
\beq M_{kj}(t) =  E(t) \eulere^{-\imagi(\varepsilon_j-\varepsilon_k) t}  \bar T_{kj}. \label{eq:matrixMkj} \eeq
If we start from the many-body ground state at $t=0$ (where the field $E(t)$ is not yet switched on), the initial condition is $b_0(0) = 1$ and all other $b_k(0)$ zero.
Equation \eqref{fig:eomforbvectormatrixformmain} is a set of coupled ordinary differential equations of first order, which can be solved numerically by standard methods. The feasibility and computational demand depends on the rank $n$ and on the sparseness of  $\mathbf{M}(t)$.

Harmonic spectra are calculated as $ Y(\Omega) = |\mathrm{FFT} [\mathrm{hann}(t) \ddot{d}(t)] |^2 $ where FFT denotes the fast-Fourier transform from the  time  to the  frequency domain  $\Omega$, hann$(t)$ is the Hann window  to increase the dynamic range, and $\ddot{d}$ is the dipole acceleration, which is proportional to the second derivative of the position expectation value
\begin{align}
x(t) &= \boldsymbol{\psi}^\dagger(t) \mathbf{h}  \boldsymbol{\psi}(t) =  \sum_{kj} b^*_k(t) b_j(t)    \bar T_{kj} = \mathbf{b}^\dagger(t)\, \bar{\mathbf{T}} \,\mathbf{b}(t) \label{eq:posexpectinbsmain}
\end{align}
where $\bar T_{kj}$ are the matrix elements of the transition matrix $\bar{\mathbf{T}}$.

\section{Results} \label{sec:results}

\begin{figure}
\includegraphics[width=0.9\columnwidth]{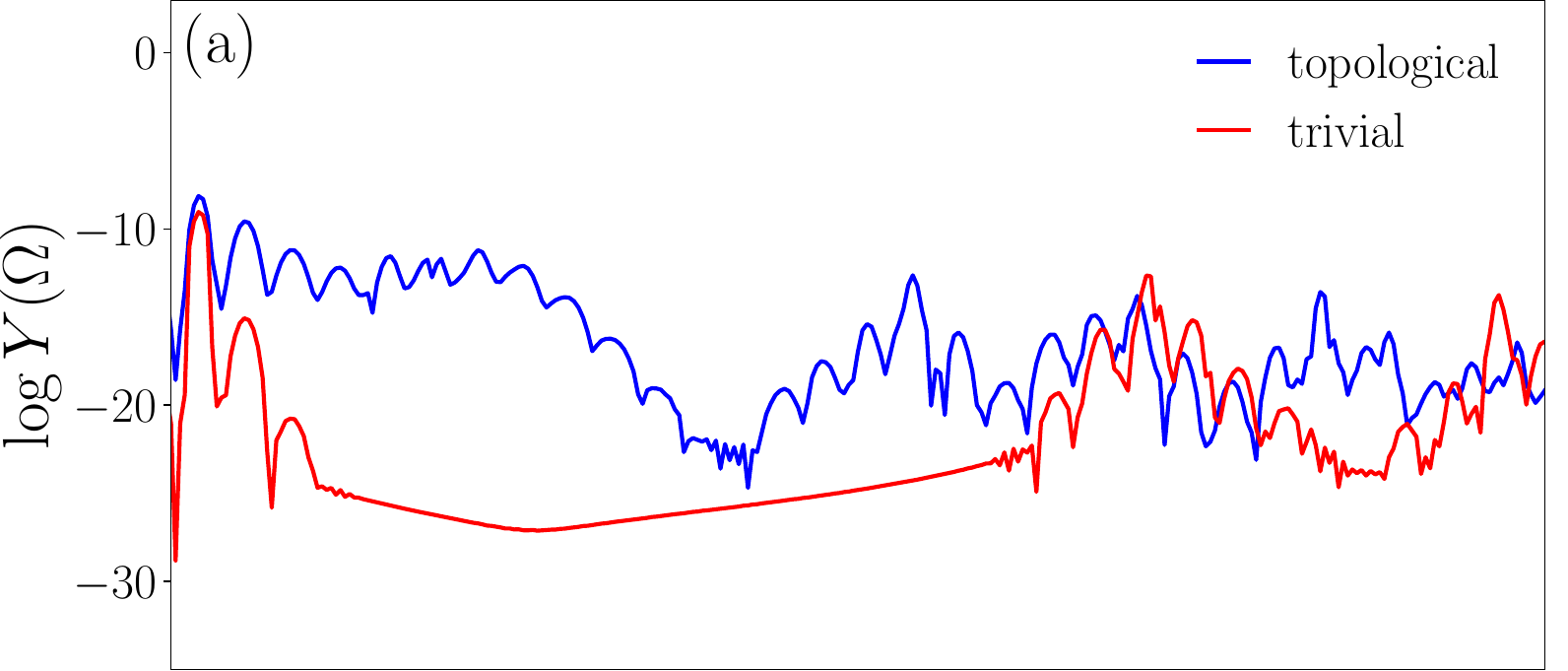}

\includegraphics[width=0.9\columnwidth]{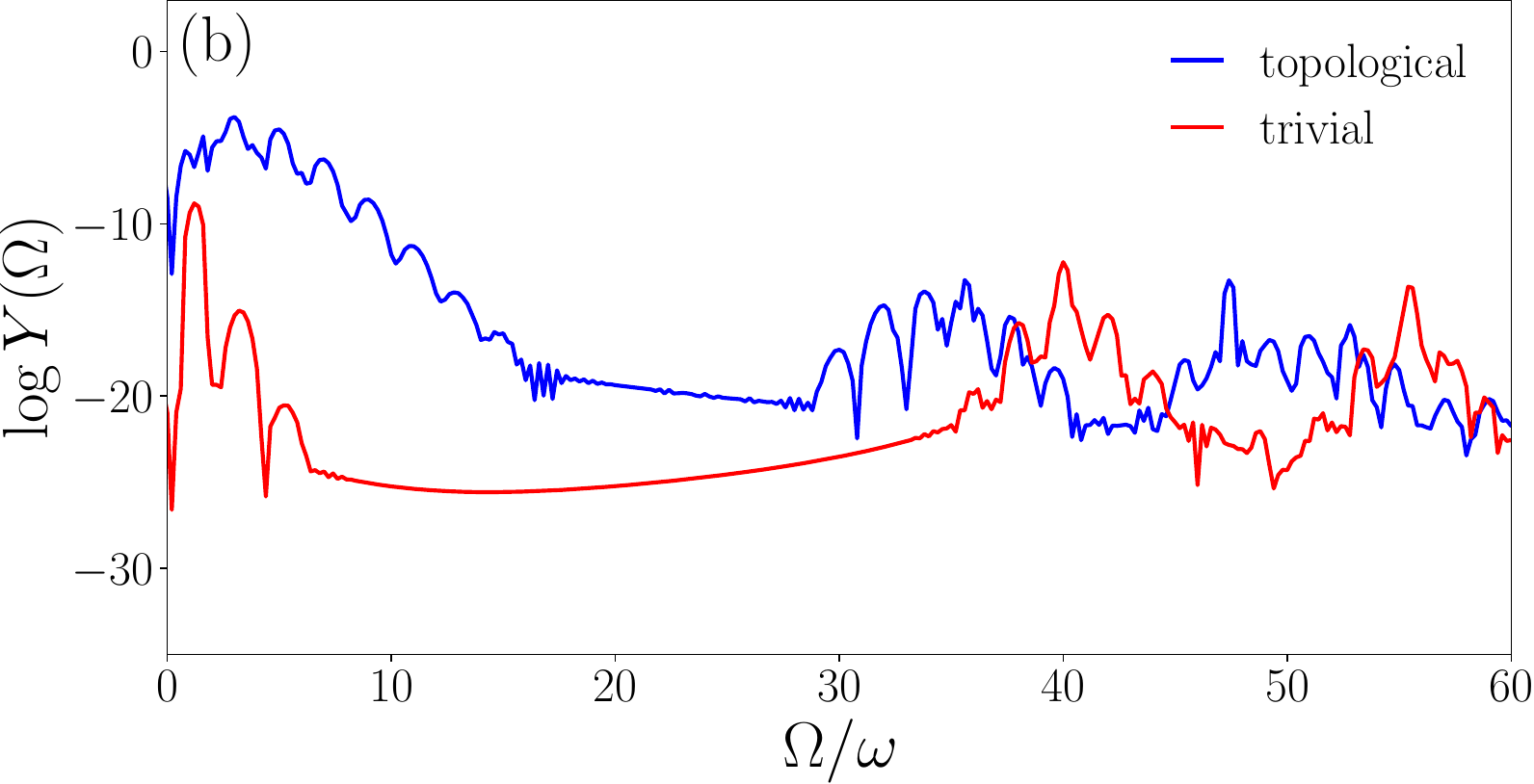}
\caption{Logarithm of the harmonic yield $\log Y(\Omega)$ vs the harmonic order $\Omega/\omega$ for (a) $U=0.1$ and (b) $U=0$. Note the huge difference between the topological and the trivial phase in the yield up to harmonic order $\simeq 20$. } \label{fig:harmonicsspec1}
\end{figure}

Figure \ref{fig:harmonicsspec1} shows harmonic spectra for the $N=12$ chain for the hopping parameters (chosen to be the same as in \cite{jurss2019high}) $v=0.18268$, $w =  0.10026$ ($v>w$, trivial) and $w=0.18268$, $v =  0.10026$ ($w > v$, topological). The Hubbard interaction was $U=0.1$ in Fig.\ \ref{fig:harmonicsspec1}(a) and $U=0$ in (b). The laser pulse was of $\sin^2$-shape,
\beq E(t) = E_0 \sin^2\left( \frac{\omega t}{2 n_\mathrm{cyc}} \right) \cos\omega t, \qquad 0 < t < n_\mathrm{cyc} 2\pi/\omega, \label{eq:pulse}\eeq
with laser frequency $\omega = 0.0049$ (i.e., wavelength $\lambda=9.3\,\mu$m), electric field amplitude $E_0 = 0.4 \omega$ (i.e., laser intensity $13.5\times 10^{10}$\,W/cm$^2$), and number of laser cycles in the pulse $n_\mathrm{cyc}=5$. The laser intensity in this work is a hundred times higher than in \cite{jurss2019high} where much longer SSH chains were studied. With the same laser intensity as in \cite{jurss2019high}, no harmonic plateau is observed for the short chains we  investigate in this work.

For the trivial case $v>w$, there is an exponential drop in the yield over the first few odd harmonics. This is very similar to the spectra for the topologically trivial phase in \cite{jurss2019high}.  The next group of harmonics is centered around harmonic order $\simeq 40$.  \makegreen{The level scheme of our system will be discussed in more detail in Sec.\ \ref{sec:few_states} below but we can pre-empt here already that harmonic order $\simeq 40$ corresponds to the energy difference between the lowest many-electron excited state to which transitions are allowed and the many-electron ground state.}    This energy difference shifts with increasing $U$ to higher energies. Other groups of harmonics at higher harmonic orders are located around energy differences that correspond to dipole-allowed transitions between higher excited many-electron states and the many-electron ground state. 

For the topological case $w>v$, the first excited many-body state to which dipole transitions are allowed is at lower energy. Hence the first group of harmonics occurs for harmonic orders $< 20$, which explains the huge difference in the harmonic yield there. In the trivial case $v>w$ there are simply no states available that could give rise to harmonic generation at such low photon energies. 

\makegreen{This explanation of the difference in the harmonic yields in terms of many-electron states is simple. What might be surprising though is the fact that the ``plateau''-like structure for harmonic orders $< 20$ in the  topological phase for such short $N=12$ chains looks already very similar to the one for the much longer chains in \cite{jurss2019high}. For the long chains with non-interacting electrons one usually employs very successfully the band structure of the corresponding bulk system to analyze the spectral features in  harmonic generation. There, two edge states appear in the band gap for the topological phase, one of which is populated in the ground state configuration. It was shown in \cite{bauer2018high, jurss2019high} that this presence of edge states  explains the high yield for below-band-gap harmonics in the topological phase. In this work, we study short chains at half filling with e-e interaction, for which a band structure does not exist.    The first few exponentially dropping harmonics solely originate from time-dependent populations of many-electron states, not from intraband harmonic generation, as there are no bands. Higher-order harmonics are due to transitions between many-electron states instead of interband transitions. In other words, the interpretation in terms of  non-interacting electrons  in bands and subsequent integration over the Brillouin zone  is replaced in the  many-electron picture by transitions between many-body states and populations thereof.}

\subsection{Dependence of HHG spectra on interaction $U$} \label{sec:dep_on_u}
\begin{figure*}

\includegraphics[width=0.49\textwidth]{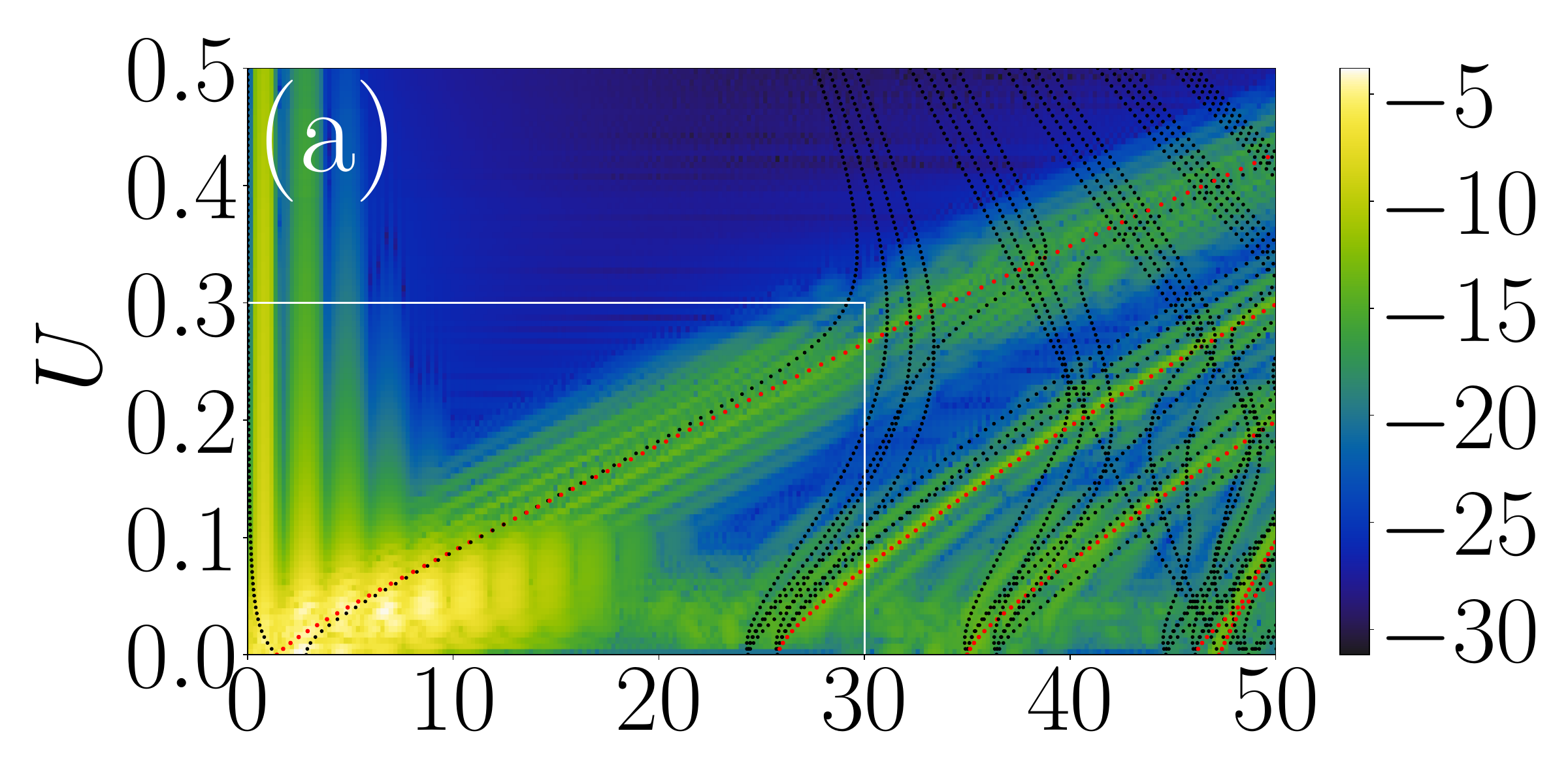}
\includegraphics[width=0.49\textwidth]{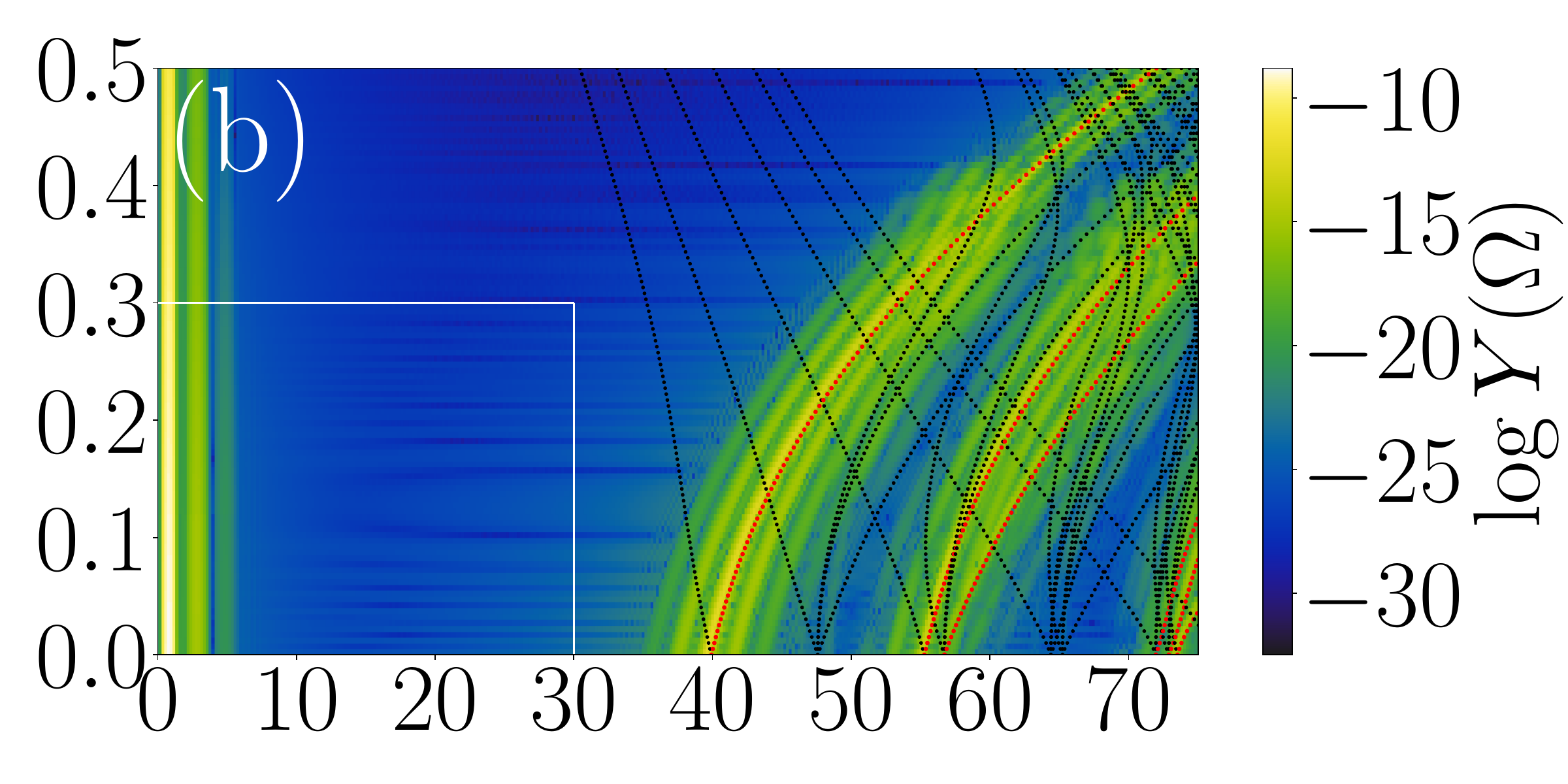}

\vspace{-0.55em}
\includegraphics[width=0.49\textwidth]{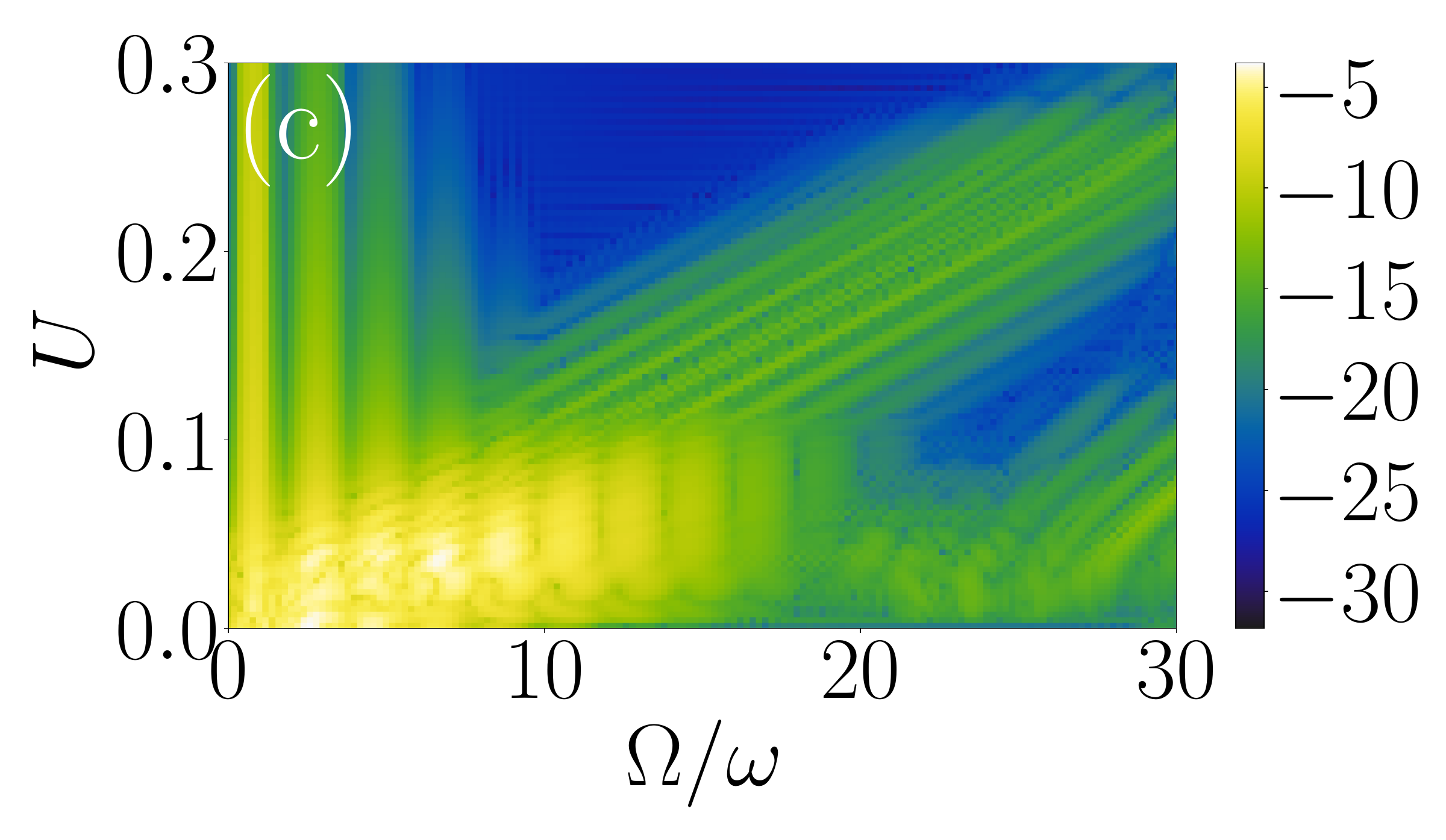}
\includegraphics[width=0.49\textwidth]{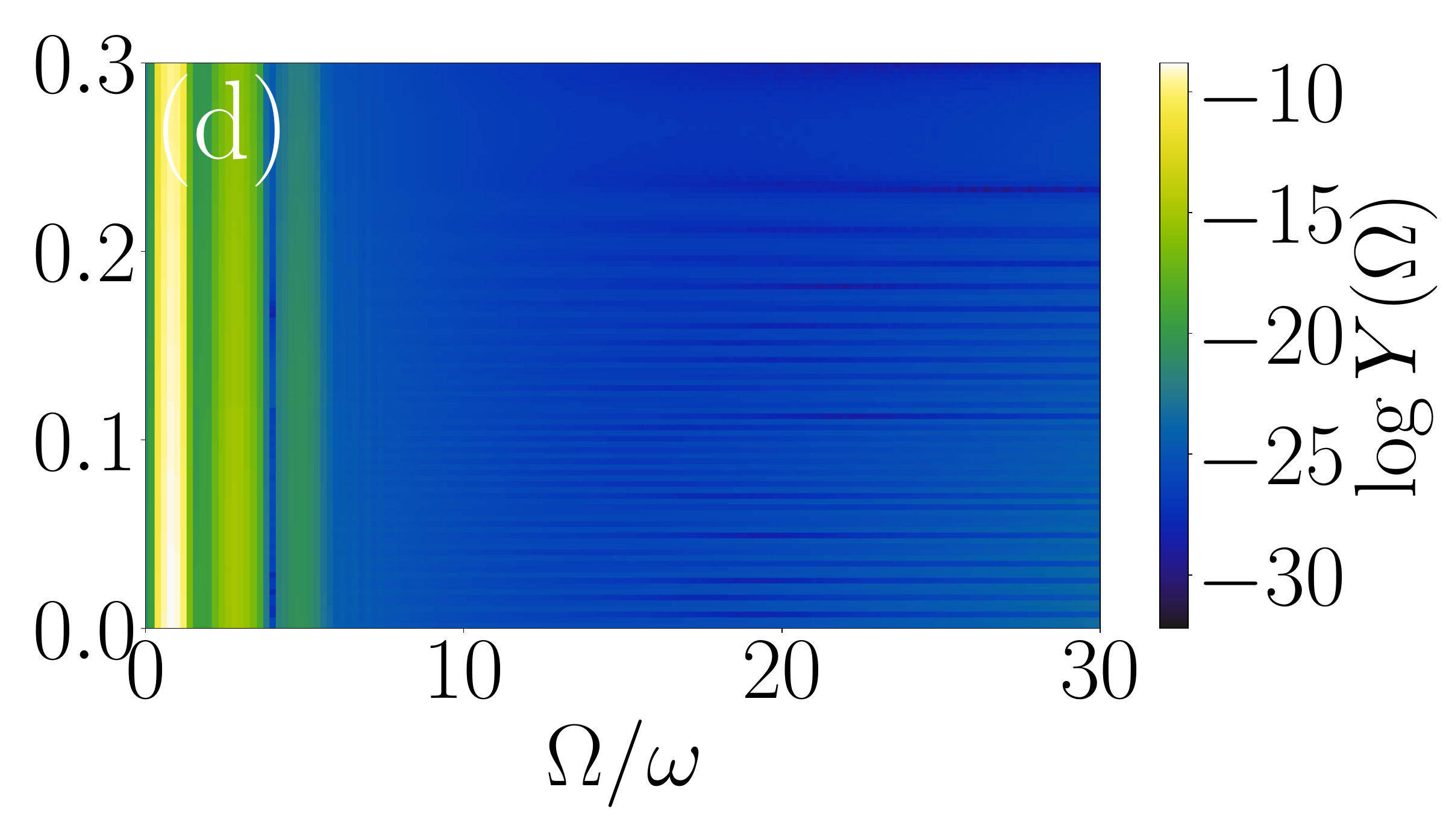}

\caption{Contour plots of the harmonic yield $\log Y(\Omega)$ vs the harmonic order $\Omega/\omega$ and the Hubbard interaction $U$ for (a) the topological and (b) the trivial cases. The energy differences between excited many-body states and the many-body ground state in units of the laser frequency are indicated by dots. Red dots indicate dipole-allowed transitions, black dots forbidden transitions.  Panels (c) and (d) zoom into the regions of the spectra indicated by the white rectangle in (a) and (b), respectively. A clear  HHG plateau up to harmonic order $\simeq 15$ is observed in the topological phase [see panel (c)] for $U<0.1$.  }
\label{fig:deponU}
\end{figure*}

Figure \ref{fig:deponU} shows HHG spectra as a function of e-e interaction $U$ for the trivial and the topological case. In both cases one sees that with increasing $U$, the emission of high-order harmonics shifts towards higher orders but with lower yield.  This is expected, because in the Mott-insulator limit of large $U$ the electrons are in the energetically most favorable configuration $\uparrow\downarrow\uparrow\downarrow\uparrow\downarrow\uparrow\downarrow\uparrow\downarrow\uparrow\downarrow$ (or $\downarrow\uparrow\downarrow\uparrow\downarrow\uparrow\downarrow\uparrow\downarrow\uparrow\downarrow\uparrow$) and cannot be polarized by the laser because electrons would have to hop to an already populated site, which is unlikely for $U\gg v,w$.  In the following, we therefore restrict ourselves to small enough $U$ such that a HHG plateau exists in the topological case. Of course, the existence of a plateau also depends on the laser parameters. In this work, we fix laser intensity, wavelength, and pulse shape to the values given after equation  \eqref{eq:pulse}.

\begin{figure}
\includegraphics[width=1\columnwidth]{fig4.pdf}

\caption{The lowest-lying relevant levels sufficient to reproduce the HHG spectra up to harmonic order $\simeq 34$ for $U=0.1$ in (a) the topological case, and (b) in the trivial case.  Here, $0, 2, 3, \ldots$ represent the states $\mathbf{u}_0, \mathbf{u}_2, \mathbf{u}_3, \ldots$. Double-headed arrows indicate allowed transitions. \makegreen{Blue arrows indicate transitions responsible for the efficient generation of harmonics in Fig.\ \ref{fig:reduced_spectra}. } The little red arrows indicate the energy of the photons in the  incoming laser pulse.}
\label{fig:important_config}
\end{figure}

\subsection{Sufficiency of only a few many-electron states}\label{sec:few_states}

\begin{figure}
\includegraphics[width=0.9\columnwidth]{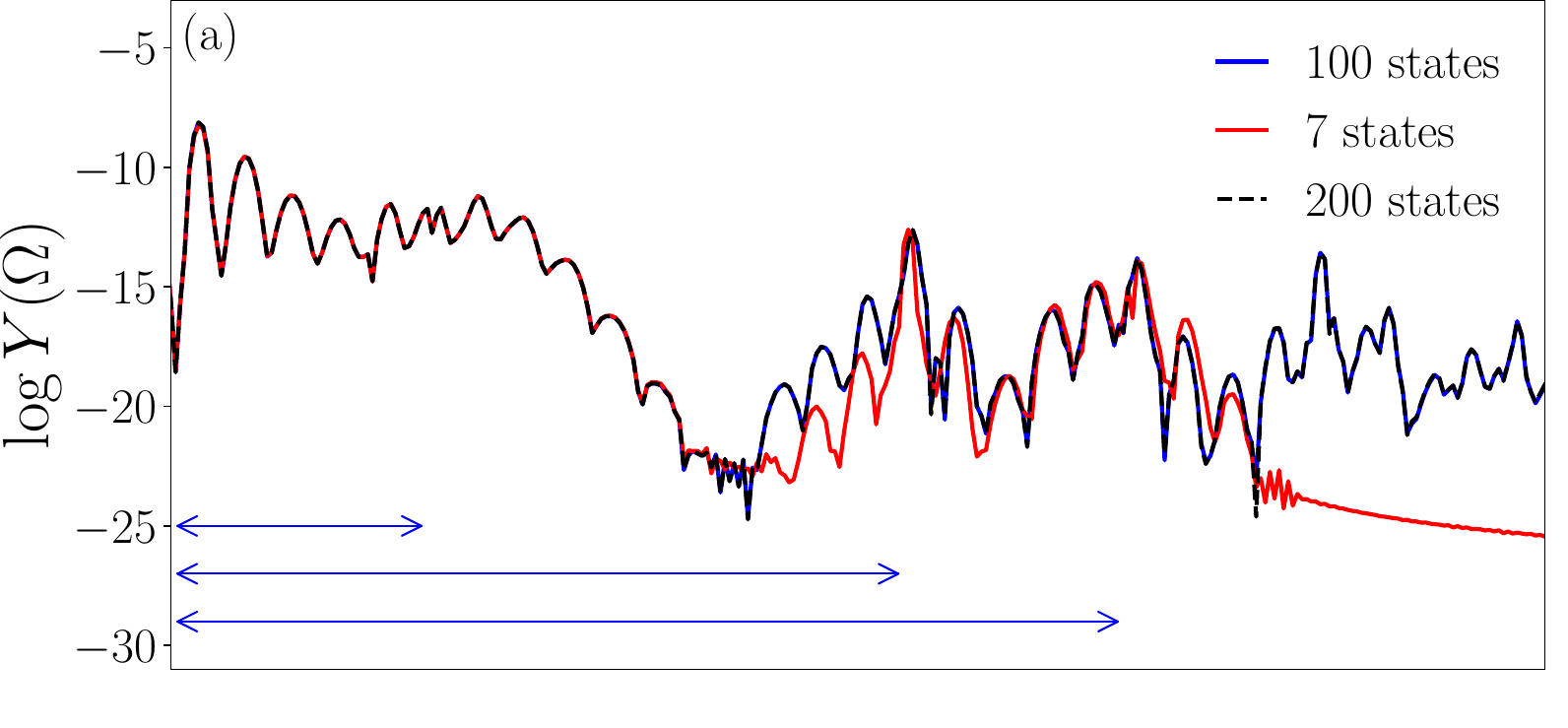}
\includegraphics[width=0.9\columnwidth]{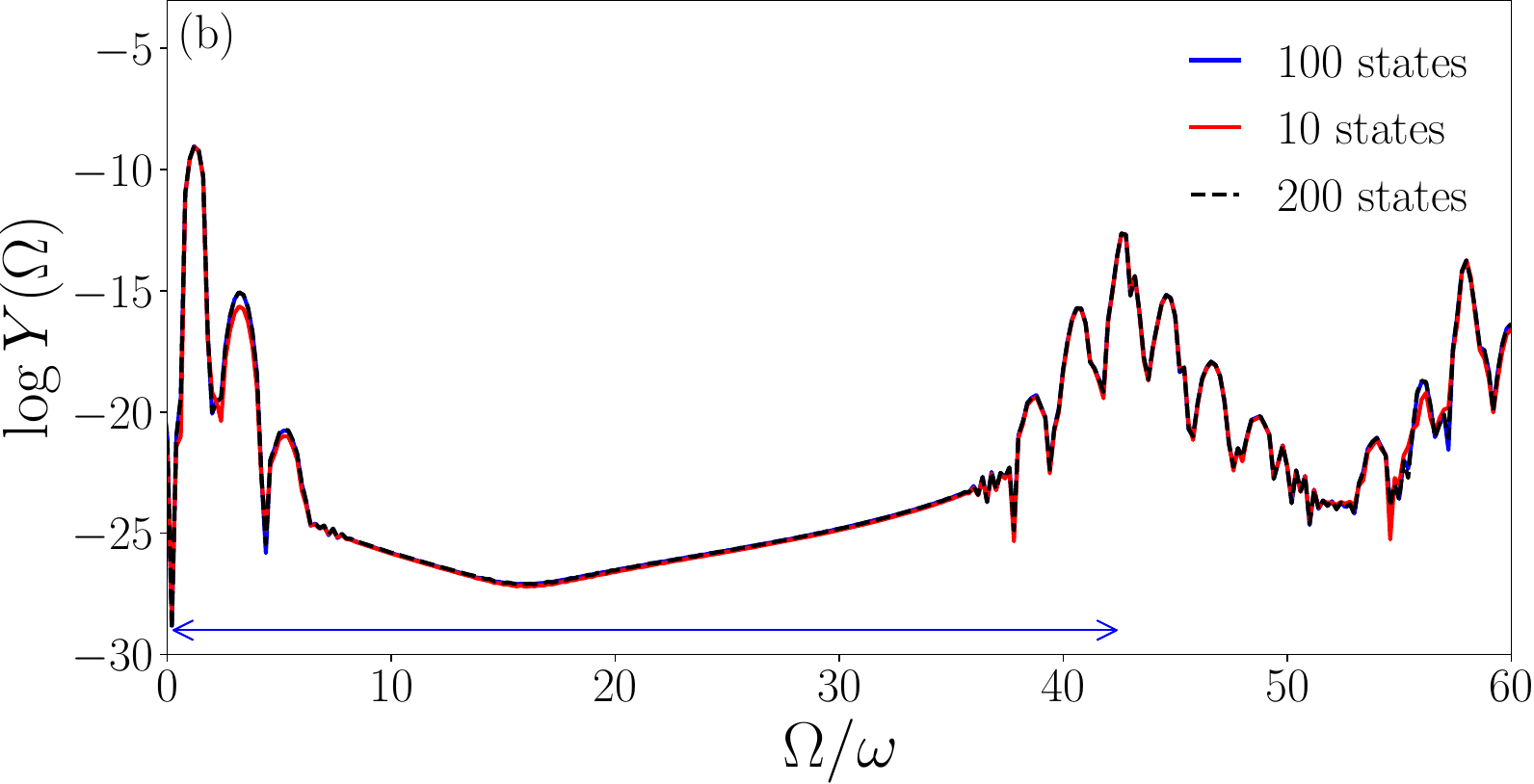}
\caption{Logarithm of the harmonic yield $\log Y(\Omega)$ as a function of the harmonic order $\Omega/\omega$ for $U=0.1$ for (a) topological case and (b) trivial case. \makegreen{ Calculations with the $100$ ($200$) lowest-lying eigenstates are shown in blue (black dashed). The results for $100$ and $200$ states lie on top of each other so that it appears safe to conclude that the spectra (up to harmonic order $60$) are converged. } Calculations with a reduced number of states ($7$ and $10$ states in (a) and (b), respectively) are shown in red. This very small number of many-body states is sufficient to reproduce harmonics up to order $\simeq 25$, covering the first plateau in (a). \makegreen{The blue horizontal arrows correspond to the blue-colored transitions in Fig.\ \ref{fig:important_config}.} }
\label{fig:reduced_spectra}
\end{figure}

In the case of twelve sites and half filling, the matrix $\mathbf{H}_0$ in \eqref{eq:H0matrixequ} is of rank $n=853776$ such that we have $853776$ eigenvalues and eigenvectors. \makegreen{We used Python and the SLEPc package (``Scalable Library for Eigenvalue Problem Computations'') to find the desired number of lowest eigenvalues and eigenstates.} Starting from the many-body ground state $\mathbf{u}_0$, not all the $853775$ excited many-body  states will be populated by the laser. For many of the excited states, the transition matrix elements $\bar T_{kj}$ are zero, which constitutes selection rules for transitions between many-body states. \makegreen{The energy differences related to these  transitions, in units of the incident laser's frequency,  are indicated in Figs.\ \ref{fig:deponU}a,b by dots, where red dots are used for allowed transitions and black dots for forbidden transitions.
Moreover, transitions, which are in principle allowed by $\bar{\mathbf{T}}$, are just unlikely, given our very small laser frequency, which is much lower than the transition to the first excited many-body state. As a consequence, we should obtain the same HHG spectra with a much reduced matrix $\mathbf{M}(t)$ in
\eqref{fig:eomforbvectormatrixformmain}, taking only a few lowest $b_k(t)$ into account and excluding those $k$ to which transitions are forbidden.}  The number of states necessary to reproduce the  HHG spectrum obtained with all states taken into account depends on the highest harmonic order of interest. 
For instance, if we are satisfied with reproducing very well only the first plateau for the topological case in Fig.\  \ref{fig:harmonicsspec1}(a) it is sufficient to take only $7$ many-electron states into account, namely $k=0,2,3,5,11,16,18$. The allowed transitions are between states $(0, 2), (0, 11), (0, 18),$ $(2, 3), (2, 5), (2, 16),$ $(3, 11), (3, 18), (5, 18), (11, 16)$, and $(16, 18)$. Figure~\ref{fig:important_config}(a) shows these levels and transitions for the topological case and $U=0.1$. \makegreen{ In this case the many-electron states $\mathbf{u_2},\mathbf{u_3}$ become almost degenerate many-body edge states discussed below in Sec.\  \ref{sec:twelve_many_body_states}. The degeneracy is up to the third digit after the comma in energy.}    
Figure \ref{fig:reduced_spectra}(a) shows that the first plateau is indeed well covered by just these seven many-body levels. 
For the trivial case, the harmonics in the same range are reproduced  with the ten  states $k=0,2,4,11,13,15,17,29,30,34$ and the transitions $(0, 2), (0, 11),  (0, 13), (0, 29), (0, 30), (0, 34), (2, 4), (2, 15),$   $(2, 17),(4, 11), (4, 13), (4, 29), (4, 30), (4, 34), (11, 15),$ $ (13, 15), (13, 17),(15, 29), (15, 30), (15, 34), (17, 29), $ $(17, 30)$ and $(17, 34)$, as shown in Fig.\ \ref{fig:reduced_spectra}(b). The states and transitions for the trivial case are depicted  in Fig.~\ref{fig:important_config}(b). The main difference compared to the topological case  in (a) is that the energy difference between ground and first accessible excited state is much larger, and that the precursor of  degenerate edge states in (a)  (i.e., $\mathbf{u_2},\mathbf{u_3}$) is absent. It might be surprising that more states are necessary in the trivial case despite the fact that there is no plateau but only exponentially dropping Brunel harmonics. The reason is that, in order to get the yield of  these low-order Brunel harmonics right down to $10^{-25}$ in dynamic range, one needs the accurate population of the ground state and thus the  excited states to which population may go. \makegreen{In the topological case, the Brunel harmonics are just not visible down to $10^{-25}$ because the plateau with harmonics  due to the first dipole-allowed  transition dominates the spectrum already at lower harmonic orders.}    

Note that by identifying the lowest relevant many-body states, the calculation of HHG in a correlated many-electron system has become computationally cheap because it reduces to solving a few-level time-dependent Schr\"odinger equation. \makegreen{ However, the numerically demanding task to calculate these relevant many-body states $\mathbf{u}_j$ in the Schr\"odinger equation \eqref{eq:H0matrixequ} in the first place remains. Moreover, the longer the chains, the closer together lie the eigenenergies $\varepsilon_j$, and the more states have to be considered in the reduced matrix $\mathbf{M}$ for the calculation of HHG. }

\subsection{Dominant configurations in the  low-lying many-body states} \label{sec:twelve_many_body_states}

We investigated the dominant configurations in the  lowest many-body eigenvectors $\mathbf{u}_j$   for $U=0.1$ and $U=0.2$ in both the trivial and topological phases of a twelve-site half-filled chain. Note that 
$U=0.1$ is similar to the value for the smaller hopping amplitude $0.10026$ ($=v$ in the topological and $w$ in the trivial case). Hence sacrificing one of these subdominant hoppings is energetically similarly  expensive as paying one $U$ for a doubly occupied site, i.e., a doublon. The situation is different for $U=0.2$ where paying  a $U$ for a doublon is more expensive than sacrificing even  a dominant  hopping.

The configuration $\downarrow\uparrow\downarrow\uparrow\downarrow\uparrow\downarrow\uparrow\downarrow\uparrow\downarrow\uparrow$ (along with its spin or space-inverted counterpart $\uparrow\downarrow\uparrow\downarrow\uparrow\downarrow\uparrow\downarrow\uparrow\downarrow\uparrow\downarrow$) is the most energetically favorable and dominates in both the ground state $\mathbf{u}_0$ and the first excited state $\mathbf{u}_1$, across both phases and for both values of $U$. 
This outcome is anticipated because it maximizes the number of possible hoppings (twelve possible intracell $v$-hoppings and ten possible intercell $w$-hoppings), thereby minimizing the kinetic energy. Additionally, since no two electrons occupy the same site, there is no energy increase from the e-e interaction $U$. However, as $\downarrow\uparrow\downarrow\uparrow\downarrow\uparrow\downarrow\uparrow\downarrow\uparrow\downarrow\uparrow$ is not an eigenstate of the Hamiltonian \eqref{eq:Hmain} other configurations mix into $\mathbf{u}_0$. \makegreen{ In fact, the differences in weights from one configuration to the next dominant one are typically less than an order of magnitude.   The next dominant ones after $\downarrow\uparrow\downarrow\uparrow\downarrow\uparrow\downarrow\uparrow\downarrow\uparrow\downarrow\uparrow$ are those with two pairs of adjacent equal-spin electrons within unit cells (in the topological case) and across unit cells (in the trivial case).} These configurations still allow for all ten $w$-hoppings (in the topological case) and all twelve $v$-hoppings (in the trivial case). The first configurations admixed into $\mathbf{u}_0$  that involve a doublon  and a holon are of the kind  
$\downarrow\uparrow\downarrow\uparrow\downarrow\updownarrow\emptysite\uparrow\downarrow\uparrow\downarrow\uparrow$ in the topological case.
Here, $\updownarrow$ indicates a doublon (i.e., a doubly occupied site) and \ $\emptysite$ \ a holon (i.e., a vacant site). Here and in the following we show in each case only one of the equivalent spin or space-inverted configurations. For $U=0.1$ the  configuration $\downarrow\uparrow\downarrow\uparrow\downarrow\updownarrow\emptysite\uparrow\downarrow\uparrow\downarrow\uparrow$
has the fourth highest weight in $\mathbf{u}_0$, where the weight is measured as the modulus square of the respective component in the eigenvector $\mathbf{u}_0$. For $U=0.2$  it has only the sixth highest weight because a doublon is energetically more expensive. 
In the trivial case, states with the doublon-holon pair at the edge of the kind $\updownarrow\emptysite\downarrow\uparrow\downarrow\uparrow\downarrow\uparrow\downarrow\uparrow\downarrow\uparrow$ have the eighth highest weight for $U=0.1$ in $\mathbf{u}_0$  but are not among the top ten dominant configurations for $U=0.2$.

As we know from Fig.\ \ref{fig:important_config} that $\mathbf{u}_2$ is the first many-body state to which a dipole-allowed transition can occur it is interesting to look at the  dominant configurations in $\mathbf{u}_2$. In the topological case, these are for both $U$ values many-body edge states with a doublon at one edge and a holon at the other, e.g.,  $\emptysite\uparrow\downarrow\uparrow\downarrow\uparrow\downarrow\uparrow\downarrow\uparrow\downarrow\updownarrow$. The situation is very different in the trivial case where the dominant configurations in $\mathbf{u}_2$ are of the kind $\downarrow\uparrow\updownarrow\emptysite\updownarrow\emptysite\updownarrow\emptysite\updownarrow\emptysite\downarrow\uparrow$ for $U=0.1$, i.e., several adjacent  doublon-holon pairs in the bulk. For $U=0.2$, in the trivial case,  the dominant configurations in $\mathbf{u}_2$ avoid doublon-holon pairs altogether but are of the type  $\downarrow\uparrow\downarrow\uparrow\downarrow\uparrow\uparrow\downarrow\uparrow\downarrow\uparrow\downarrow$.

\makegreen{A future goal is to develop spectroscopic methods that allow to measure correlation functions of doublons and holons, e.g.,  at the edges. The HHG yield studied in this work is only sensitive to the electron density, the level scheme, and the transition dipole matrix elements, and thus only indirectly to the electronic configurations.}

\section{Conclusion and Outlook} \label{sec:summary}
In this work, we studied the effect of e-e interaction on harmonic generation in half-filled  SSH-Hubbard chains. For non-interacting electrons, long but finite SSH chains  show the well-known topological edge states that were found to play a significant role in the enhancement of the harmonic yield in the topological phase. 
Since we used the method of exact diagonalization for the interacting SSH-Hubbard chains, we were restricted to short chains, no longer than twelve sites. However, these short chains already  showed the precursors of what would be topological edge states for longer chains without interaction. 

There are two main results of this work:  (i) The huge difference in the harmonic yield between trivial and topological SSH phase observed earlier survives e-e interaction (unless the interaction is so large that harmonic generation ceases to exist anyway), and (ii) a few low-lying many-electron states are sufficient to describe harmonic generation in the small finite chains considered, including non-perturbative plateaus. With increasing system size, more and more of such many-electron states will contribute. 

Future work is planned to be devoted to a time-dependent (multi-configurational) Hartree-Fock treatment of the same chains in order to investigate the role of many-body correlation, and to 2D topological systems with interaction such that the polarization properties of the harmonics can be investigated. Further, the field of topological invariants and robustness for interacting systems is still in its infancy. Hence it will be of general interest to study how topological features   behave as the interaction between particles increases. Harmonic generation is a particularly attractive observable because it is purely optical and non-invasive. Using additional laser fields, one may dress the system to become Floquet-topological, and thus make topological features tunable.

\section*{Acknowledgment}

	    Funding by the German Research Foundation - SFB 1477 ``Light-Matter Interactions at Interfaces,'' Project No. 441234705, is gratefully acknowledged.

\bibliographystyle{apsrev4-2}

\bibliography{Hubbard_references}

\end{document}